# Category Theoretic Analysis of Photon-based Decision Making

Makoto Naruse

*Network System Research Institute, National Institute of Information and Communications Technology,*
*4-2-1 Nukui-kita, Koganei, Tokyo 184-8795, Japan*
*Université Grenoble Alpes, CNRS, Institut Néel, 38000 Grenoble, France Grenoble*
naruse@nict.go.jp

Song-Ju Kim

*Graduate School of Media and Governance, Keio University,*
*5322 Endo, Fujisawa, Kanagawa 252-0882, Japan*
songju@sfc.keio.ac.jp

Masashi Aono

*Faculty of Environment and Information Studies, Keio University,*
*5322 Endo, Fujisawa, Kanagawa 252-0882, Japan*
aono@sfc.keio.ac.jp

Martin Berthel, Aurélien Drezet, Serge Huant

*Université Grenoble Alpes, CNRS, Institut Néel, 38000 Grenoble, France*
martin.berthel@u-bordeaux.fr, aurelien.drezet@neel.cnrs.fr, serge.huant@neel.cnrs.fr

Hirokazu Hori

*Interdisciplinary Graduate School, University of Yamanashi, Takeda, Kofu,*
*Yamanashi 400-8511, Japan*
hirohori@yamanashi.ac.jp

Decision making is a vital function in this age of machine learning and artificial intelligence, yet its physical realization and theoretical fundamentals are still not completely understood. In our former study, we demonstrated that single-photons can be used to make decisions in uncertain, dynamically changing environments. The two-armed bandit problem was successfully solved using the dual probabilistic and particle attributes of single photons. In this study, we present a category theoretic modeling and analysis of single-photon-based decision making, including a quantitative analysis that is in agreement with the experimental results. A category theoretic model reveals the complex interdependencies of subject matter entities in a simplified manner, even in dynamically changing environments. In particular, the octahedral and braid structures in triangulated categories provide a better understanding and quantitative metrics of the underlying mechanisms of a single-photon decision maker. This study provides both insight and a foundation for analyzing more complex and uncertain problems, to further machine learning and artificial intelligence.

*Keywords*: Decision making; category theory; single photon; machine learning; multi-armed bandit problem; system modeling.

## 1. Introduction

To maximize the total reward sum from multiple slot machines, we need to find a machine that provides the highest reward probability. However, too much exploration may result in excessive losses, as the definition of a best machine may change over time. Conversely, a quick decision, or insufficient exploration, might result in overlooking the best machine. This is called the *exploration–exploitation dilemma* tradeoff, formulated as the multi-armed bandit problem (MAB) in the literature of machine learning.[1,2,3] The MAB problem is important for various practical applications, such as information network management,[4,5] web advertisements,[6] Monte Carlo tree searches,[7] and clinical trials.[8,9] Meanwhile, decision theories have been intensively investigated using the quantum mechanical notions for modeling human decision making.[10-12]

For vital applications, it is important to develop high-performance MAB algorithms that can be executed on conventional electronic computing platforms. Examples of these algorithms include softmax,[1,2] upper-confidence bound,[13] and tug-of-war.[14,15] Meanwhile, we have been investigating *physical* principles and technologies, for realizing *artificially constructed physical decision-making mechanisms*. One motivation behind this research is to deepen the understanding of intelligent abilities inherent in natural phenomena.[16] The aim is to extract and utilize these, to pave the way for developing novel intelligent devices, using state-of-the art materials and/or photonics technologies.[17,18] In our former study,[3] we proposed an architecture in which the quantum attributes of a *single photon* are utilized for decision making. We demonstrated (experimentally) accurate and adaptive decision making, using the nitrogen-vacancy (NV) center in a nanodiamond as the single-photon source. Thanks to the quantum nature of light, single-photon detection is both immediately and directly associated with decision making, which is a decisive step toward achieving an autonomous intelligent machine based on a purely physical mechanism.

In this study, we investigate the theoretical background for clarifying the underlying mechanisms of a single-photon decision maker. In Ref. 3, the polarizations of single photons were adaptively configured, such that a higher-reward-probability slot machine was selected. Using the geometry-based modeling described in this study, we can determine the dynamic change of polarizations. Based on this study, we discuss an abstracted model of the single-photon decision maker via the notions of *category theory*.[19-24] Category theory is a branch of mathematics that formalizes mathematical structures into collections of objects and morphisms. Category theory extracts the essence of all mathematical subjects, including a dynamically changing environment, to reveal and formalize simple, powerful patterns of thinking. In this study, we analyze how decision making is conducted using a category theoretic picture. The complex interdependencies involved in decision making problems are clarified by category theory. In particular, we show that the octahedral and braid structures (in triangulated categories) provide a qualitative understanding, and quantitative metrics, of the underlying mechanisms of a single-photon decision maker.

The study is organized as follows. In Sec. 2, we review single-photon-based decision maker experiments.[3] Section 3 discusses geometry-based modeling and analysis, including experimental data analysis. Section 4 provides the structural analysis of decision making, based on topological features. Section 5 presents a category theoretic modeling and analysis of a photon decision maker. Numerical characterizations are also discussed, to investigate the correlation between category theory modeling and an actual physical system. Section 6 concludes the study.

## 2. Single-Photon-Based Decision Making

For the simplest case that preserves the essence of the MAB problem, we considered a player who selects one of two slot machines (slot machine 1 or slot machine 2) with the goal of maximizing the reward. A polarizing beam splitter (PBS) was prepared, as shown in Fig. 1. Initially, the linear polarization of the single-photon input was orientated at $\pi/4$ (with respect to the horizontal), enabling the photon to be detected by Channel 1 (Ch.1) or Channel 2 (Ch.2) with a 50:50 probability. However, the total probability of photon detection, by either Ch.1 or Ch.2, is unity. Here, photon detection by Ch.1 or Ch.2 is immediately associated with the decision to select slot machine 1 or 2, respectively. This is a notable aspect of the single-photon decision maker, in the sense that the dual probabilistic (wave) and particle attributes of a single photon are both utilized. A single photon, polarized at an orientation of $\pi/4$ (with respect to the horizontal), indicates that the system is making a thorough search for a better machine. From this initial condition, polarization is reconfigured using the following strategy:

> **[Polarization updating strategy in the single-photon decision maker]**
>
> If the selected machine successfully dispenses a reward, polarization is shifted towards the selected machine. If no reward is dispensed from the selected machine, polarization is moved in the direction of the non-selected machine. By iterating this process, the system guides us to a decision in which we select the correct solution, meaning that the higher-reward-probability machine is selected. The orientation of polarization is quantified by a polarization adjuster value, described in detail in Ref. [3].

Through theoretical modeling and analysis, this study aims to clarify the underlying mechanisms, then explain why the single-photon decision maker (described above) can derive the correct decision.

Here, we briefly review the experimental system and results of single-photon-based decision making, and conduct an analysis of the experimental data. An individual photon was supplied by a single NV color center[25] in a surface-purified 80-nm nanodiamond.[26,27] It passed through a polarizer, and a zero-order half-wave plate, and impinged on the PBS. Two avalanche photodiodes (used for photon detection) were connected to a time-correlated single-photon-counting system. Photon detection is associated with the decision to select a slot machine. Based on the pay-off results from the selected slot machine, linear polarization was configured (towards the vertical or horizontal), by rotating a half-wave plate mounted on a rotary positioner (Fig. 1).

Let the initial reward probabilities of slot machines 1 and 2 be given by $P_1 = 0.8$ and $P_2 = 0.2$, respectively, which means that selecting slot machine 1 is the correct decision. The reward probability is inverted every 150 cycles, to represent the ability to adapt to environmental changes. The decision maker repeated 600 consecutive plays, 10 times. The solid curve in Fig. 2 shows the evolution of the correct selection rate, given by calculating the number of correct decisions divided by the number of repeat cycles. This gradually increased toward unity. Because the reward probability was swapped, the correct selection rate dropped every 150 cycles, but quickly recovered. The dotted curve in Fig. 2 shows the case in which the initial reward probabilities are $P_1 = 0.6$ and $P_2 = 0.4$. Because the difference between these probabilities was less than the previous case, making accurate decisions became more difficult. Although the performance degraded relative to the previous situation, a gradual increase in the correct selection rate, and an adaptation to environmental change, were still observed.

## 3. Geometry-Based Modeling

Several probabilistic and unobservable processes are involved in a decision making problem, which complicates its comprehension. First, machine selection is probabilistic, in that the polarization of a single photon strongly affects the resulting decision, but each individual decision is probabilistic if polarization is not completely horizontal or vertical. Second, we cannot be certain that the selected machine has a relatively higher reward probability, as we only observe the dispensed reward. To clearly represent the problem, we term slot machines with relatively larger and smaller reward probabilities as "*GOOD*" and "*BAD*" machines, respectively. Note that GOOD/BAD information is a relationship that is hidden from the decision maker. Finally, the result of a single play of a slot machine is probabilistic, even if the selected machine is the GOOD machine. We denote the situation when the selected machine does, or does not dispense a reward as "*WIN*" or "*LOSE*," respectively. Thus, there are eight possible combinations with regard to {Machine selection 1/2}, {GOOD/BAD machines}, and {WIN/LOSE}. These are schematically represented as follows:

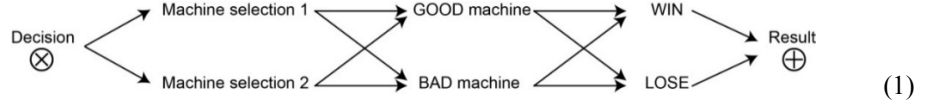

(1)

Note that each pair of alternatives in Eq. (1), i.e., {Machine selection 1/2}, {GOOD/BAD machine}, and {WIN/LOSE}, form an *orthogonal* basis with respect to a certain reference. However, they are transformed into a mixture in the subsequent pair. For example, the "WIN" state is orthogonal to "LOSE," but comprises the two cases of "GOOD machine Wins" and "BAD machine Wins." In turn, the orthogonal states, "Machine selection 1" and "Machine selection 2," are injected into "GOOD machine" and "BAD machine." This is based on hidden variables (set by the casino), from which a player is only able to observe the alternative results of {WIN/LOSE}. This viewpoint serves as an entry to the geometry-based analysis that follows, and the category theoretic paradigm.

### 3.1. *DECISION circle, WIN/LOSE circle, and their dynamics*

We quantify and analyze the problem by introducing two circular diagrams, each having a radius of unity. One of these is what we call the **"DECISION circle,"** which is concerned with the machine selection probability specified by the polarization of the single-photon decision maker. As shown in Fig. 3(a), $\phi/2$ is the angle of deviation from $\pi/4$ (ranging from $-\pi/4$ to $+\pi/4$). We regard the projection to the vertical or horizontal axis as the probabilities of selecting slot machine 1 or 2, respectively. Rigorously speaking, the square of the projection should be considered as the probability that satisfies the condition that the sum of the two projections is unity. However, for simplicity, we assume linear correspondence in this study.

The other circle denotes the WIN/LOSE probability, and is referred to as the **"WIN/LOSE circle."** Here, the GOOD machine is represented by a point on the circle that deviates from the $\pi/4$ direction by $\theta/2$, schematically shown in Fig. 3(b). The vertical and horizontal projections indicate the win and lose probabilities, respectively. The BAD machine is characterized as a point on the circle that deviates from the $\pi/4$ direction by $-\theta/2$. By assuming $\theta > 0$, the winning probability of the GOOD machine is always higher than the BAD machine, which is consistent with the definition of a GOOD/BAD machine.

Next, we consider an initial condition of the single-photon decision maker, where the polarization of a single photon is given by $\pi/4$, i.e., $\phi = 0$ (Fig. 3(c)). In this situation, the probabilities of selecting slot machine 1 or 2, denoted by the projections $P_1^{(S)}$ or $P_2^{(S)}$ to the vertical or horizontal axis, respectively, are $1/\sqrt{2}$ (based on the DECISION circle). To explain this, assume that the selected machine is slot machine 1, and is also the GOOD machine. The probability of winning should be the product of the reward probability of the GOOD machine, and its selection probability (in this case, the selection probability is $P_1^{(S)}$). By geometrically describing this situation using the WIN/LOSE circle, the winning probability is given by the projection of a vector with magnitude $P_1^{(S)}$ on the GOOD machine, schematically shown in Fig. 3(d).

Next, consider a case in which $\phi$ is slightly increased, but is still smaller than $\theta$ ($\theta > \phi > 0$) (Fig. 3(e)). As in the initial condition, the winning probability of the GOOD machine is given by the projection of a vector along the GOOD machine, with the probability of machine selection. That is, the WIN probability (when the selected machine is GOOD) is given by

$$P_{GW} = \sin\left(\frac{\pi}{4}+\frac{\phi}{2}\right)\sin\left(\frac{\pi}{4}+\frac{\theta}{2}\right), \tag{2}$$

whereas the LOSE probability (with the GOOD machine selected) is given by

$$P_{GL} = \sin\left(\frac{\pi}{4}+\frac{\phi}{2}\right)\cos\left(\frac{\pi}{4}+\frac{\theta}{2}\right), \tag{3}$$

where "GW" and "GL" are abbreviations of "Good machine Wins" and "Good machine Loses," respectively. $P_{GW}$ and $P_{GL}$ are plotted on the vertical axis of the WIN/LOSE circle. Similarly, when the selected machine is BAD, the WIN and LOSE probabilities are given by

$$P_{BW} = \cos\left(\frac{\pi}{4}+\frac{\phi}{2}\right)\sin\left(\frac{\pi}{4}-\frac{\theta}{2}\right) \quad (4)$$

and

$$P_{BL} = \cos\left(\frac{\pi}{4}+\frac{\phi}{2}\right)\cos\left(\frac{\pi}{4}-\frac{\theta}{2}\right), \quad (5)$$

respectively. The difference between Eq. (2) and Eq. (4), and the difference between Eq. (5) and Eq. (3), are respectively simplified/given by

$$P_{GW} - P_{BW} = \sin\left(\frac{\phi+\theta}{2}\right), \quad P_{BL} - P_{GL} = \sin\left(\frac{\theta-\phi}{2}\right). \quad (6)$$

Let $\theta$ be a positive fixed value, then the derivatives of Eq. (6), with respect to $\phi$, are given by

$$\frac{d(P_{GW} - P_{BW})}{d\phi} \propto \cos\left(\frac{\phi+\theta}{2}\right), \quad \frac{d(P_{BL} - P_{GL})}{d\phi} \propto -\cos\left(\frac{\theta-\phi}{2}\right), \quad (7)$$

indicating that the former and latter terms give positive and negative values, respectively, based on the condition that $\theta > \phi > 0$. In other words, the probability difference ($P_{GW} - P_{BW}$) is increasing, while $P_{BL} - P_{GL}$ is decreasing. This means that by increasing $\phi$, the probability of selecting the GOOD machine also increases (Fig. 3(f)). Based on the repetition strategy of the single-photon decision maker described earlier, $\phi$ is increased.

When $\phi = \theta$, both $P_{GL}$ and $P_{BL}$ are projected onto the same point in the WIN/LOSE circle, diminishing their difference, which is confirmed by the second equation in Eq. (6).

Next, consider the case where $\phi > \theta$ (Fig. 3(g)). Here, the difference along the LOSE axis is modified from the previous case (the second term in Eq. (6)), and is given by

$$P_{GL} - P_{BL} = \sin\left(\frac{\phi-\theta}{2}\right). \quad (8)$$

The derivative of Eq. (8), with regard to $\phi$, is proportional to $\cos[(\phi - \theta)/2]$, which is always positive, as $\phi > \theta$. This indicates that $P_{GL} - P_{BL}$ is increasing (Fig. 3(h)).

Finally, $\phi$ ultimately reaches the vertical axis, i.e., $\phi/2 = \pi/4$ (Fig. 3(i)). Here, the probability of selecting machine 1 is unity in the DECISION circle. The machine selection vector is projected directly onto the axis of the GOOD machine in the WIN/LOSE circle, whereas the projection along the axis of the BAD machine completely disappears. Therefore, the WIN/LOSE probability stems only from the property of the GOOD/BAD machines specified by $\theta$, which is schematically shown in Fig. 3(j). Using the abovementioned mechanism, the polarization $\theta$ of the single-photon decision maker is autonomously directed towards the GOOD machine.

We conducted simulations to examine the mechanism discussed above. We first considered the case in which the reward probabilities of slot machines 1 and 2 are 0.6 and 0.4, respectively. In the WIN/LOSE circle, this corresponds to the case where the GOOD machine is characterized by the value $\theta/2$, for which the vector (0.4, 0.6) intersects the unit circle. Specifically, $\theta/2$ corresponds to 0.20 rad. The BAD machine is represented in the circle as a deviation from $\pi/4$ by $-\theta/2$. Here, 500 consecutive plays were repeated 500 times. We evaluated the following conditional probabilities to examine the dynamics behind the decision making that is compatible with the theoretical modeling:

(i) The probability of winning by selecting the GOOD machine if the result is "win": $P(\text{GOOD WIN} | \text{WIN})$;
(ii) The probability of winning by selecting the BAD machine if the result is "win": $P(\text{BAD WIN} | \text{WIN})$;
(iii) The probability of losing by selecting the GOOD machine if the result is "lose": $P(\text{GOOD LOSE} | \text{LOSE})$;
(iv) The probability of losing by selecting the BAD machine if the result is "lose": $P(\text{BAD LOSE} | \text{LOSE})$.

These are depicted respectively by the solid red, dashed green, dotted blue, and dash-dot magenta curves in Fig. 4(a, i). Here, the probabilities are calculated as ensemble averages over the samples.

The difference between (i) and (ii) increases over time throughout the playing cycles, which is consistent with the increase in magnitude of the arrow projected onto the WIN axis in the WIN/LOSE circle. Conversely, the difference between (iii) and (iv) decreases until a given playing cycle is reached (approximately cycle number 9). This is consistent with the decrease in magnitude of the arrow projected onto the LOSE axis in the WIN/LOSE cycle, in which $\phi < \theta$ (Fig. 3(e)). After the cycle, the difference between (iii) and (iv) increases, which is consistent with the theoretical modeling for the case where $\phi > \theta$ (Fig. 3(g)).

Fig. 4(a, ii) shows the same analysis when the reward probabilities of slot machines 1 and 2 are 0.8 and 0.2, respectively. As the difference between the GOOD and BAD machines has increased ($\theta/2$ corresponds to 0.54 rad) it is easier for the decision maker to find the GOOD machine. The dynamics of the conditional probability follows similar trajectories to the previous case, and eventually change from $\phi < \theta$ to $\phi > \theta$, which corresponds to the intersection between $P(\text{GOOD LOSE} | \text{LOSE})$ and $P(\text{BAD LOSE} | \text{LOSE})$ appearing in cycle 11.

### 3.2. *Experimental analysis*

We analyzed the experimental data generated by single-photon-based decision making based on the formula described above. As reviewed in Sec. 2, the correct selection rate approaches unity as time elapses. We focused on the initial 150 cycles, starting from the initial condition that the polarization is $\pi/4$, or equivalently, $\phi = 0$.

When the reward probabilities of slot machines 1 and 2 were 0.8 and 0.2, the factorized evolution of conditional probabilities are summarized in Fig. 4(b, ii). Due to the number of repeated samples being limited to 10 in the experiment, which is small, the curves occasionally fluctuated. However, similar dynamics were successfully observed in Fig. 4(a, ii), as predicted by the theory.

Fig. 4(b, i) summarizes the conditional probabilities when the reward probabilities of slot machines 1 and 2 are given by 0.6 and 0.4, respectively. The blue dotted curve $[P(\text{GOOD LOSE} | \text{LOSE})]$ and dash-dot magenta curve $[P(\text{BAD LOSE} | \text{LOSE})]$ intersect each other, which is consistent with the theory (Fig. 4(a, i)). However, the difference between $[P(\text{GOOD WIN} | \text{WIN})]$ and $[P(\text{BAD WIN} | \text{WIN})]$ decreased during the initial 24 cycles, which does not agree with the abovementioned theoretical analysis. Note that the initial condition of $\phi = 0$ indicates that the single-photon decision maker is exploring both selections. Therefore, $\phi$ may occasionally take a negative value. In such cases, the probability of selecting slot machine 2 (which is on the horizontal axis of the DECISION circle) becomes larger, and the machine simulation is projected onto the BAD machine in the WIN/LOSE circle. Hence, in such situations, $P_{GW} - P_{BW}$ decreases.

### 4. Structural Analysis of Decision Making

In the previous section, we discussed the underlying structure inherent in the decision-making problem, via simple geometry-based modeling and analysis. However, the complex interdependencies involved in the subject matter have not yet been considered. For example, the polarization updating strategy tells us that the polarizer setup depends on both the current decision and the betting result, whereas the former simple model does not represent this relationship. Furthermore, the effects of environmental conditions have not been considered. These perspectives become more important when considering adaptive decision making, or autonomous intelligence in dynamically changing environments. This issue is closely related to the problem of describing, or formalizing, non-equilibrium open systems in physics. Given these considerations, in this section we further generalize single-photon-based decision making by using the notion of category theory.

Before proceeding to category theoretic descriptions, we analyze the composite system of the casino and the player (or decision maker) based on its topological structure. As the basis of structural analysis, we consider a phase space composed of the casino, player, and the environment, to describe the system.

The first step for constructing a simplified mathematical description involves identifying the apparent components of the subject matter, given by Eq. (1). These components are "Decision," "Machine 1," "Machine 2," "GOOD machine," "Bad machine," "WIN," and "LOSE." These components overlap each other, as schematically shown in Fig. 5(a). We refer to this as the covering space. By describing the components using nodes, and the relations using links, the composition can be abstracted by a diagram, as shown in Fig. 5(b). The environmental entities of the system determine the connections between the nodes. For example, the decision made by the single-photon decision maker to select "Machine 1," is represented by the link from **D** to **M1**, whereby the photon environment affects the single-photon detection. In addition, the casino determines which of the Machines (1 or 2) is to be the "GOOD" machine (the links from **M1** (or **M2**) to **G**). Furthermore, certain probabilistic attributes in the slot machine affect the "WIN" or "LOSE" state, which determines the link from **G** (or **B**) to **W** (or **L**).

Next, we focus on the casino. Suppose there were no differences between "GOOD" and "BAD" machines, meaning there were no differences in the reward probabilities. Choosing "Machine 1" or "Machine 2" would yield the same expected rewards. In this situation, the intension of the casino, which is concerned with the difference between "GOOD" and "BAD" machines, leads to a singularity, as explained below. Once the casino assigns "Machine 1" (or "Machine 2") as a "GOOD" machine, the topological symmetry in Fig. 5(b) is broken; either of the structures shown in Fig. 5(c,i) and (c,ii) emerge, where the "GOOD" machine is related to "Machine 1" and "Machine 2," respectively.

In the case of Fig. 5(c,i), the player, or the "Decision", is highly likely to be connected to the "WIN" node (**W**) by choosing "Machine 1" (**M1**), whereas in the case of Fig. 5(c,ii), the path via "Machine 2" (**M2**) yields a high likelihood of connection to the "WIN" node. This is what we call singularity (with regard to WIN or LOSE), induced by the intension of the casino. Multivaluedness appears, regarding the WIN and LOSE nodes. For example, in the case of Fig. 5(c,i), winning is more likely to be achievable via **M1** rather than **M2**, as schematically shown in Fig. 5(d,i). Conversely in the case of Fig. 5(c,ii) there is a greater chance of winning via **M2** rather than **M1** as shown in (Fig. 5(d,ii)). In other words, multivaluedness means that there can be different probabilities to reach the WIN state, via the GOOD or BAD machine. Thus, the aim of the decision-making process is to identify the singularity, and to break the initially symmetric system (Fig. 5(b)), into an asymmetric one (Fig. 5(c)). This action is taken to ensure that the path to the "GOOD" machine is taken.

In order to analyze decision making based on category theory, we consider a composite system ($P \oplus Q$) of player ($P$) and casino ($Q$), and include the entire environmental system as a monoidal category ($\mathcal{M}$). It evolves with successive decision making, slot machine playing, receipt of reward, and adjustments in decision making. Hereafter, we refer to a single procedure of these operations as a "cycle." This is described by the endomorphism ($1_{\mathcal{M}}$) of the monoidal category, to be described in the next section;

$$\begin{array}{c} P \oplus Q \\ \cup \\ M \circlearrowleft 1_M \end{array}$$

. (9)

We assume that the composite system is a type of Abelian category, because the system takes only a single path in the covering space (based on the particle nature of a single photon), and all the processes under consideration are additive.

It is essential for our decision maker that the particle features of a single photon are guaranteed to take a single path in the diagram, for each single cycle. For each cycle, the entire system evolves into a state in the covering space where succeeding cycles will be performed in renewed relations, based on the adjustment procedure of the decision-making process (which is physically the control of the optical polarizer). This leads to an accumulation of the history of the exploration in the covering space. All of the history, memorized in the polarizer setting, determines the present connection diagram. This renewal procedure corresponds to the feature of a derived category, where the long-term inheritance of cohomology plays an essential role in the evolution, following the form of a triangular structure. This structure is referred to as the triangulated category, described in Sec. 5.

It must be emphasized that the topological analysis of the covering space is essential for evaluating singularities inherent in the system, including the environment. Multivaluedness emerges, and involves different structures of the system under analysis, characterized by a certain physical quantity. We will see in the next section that the essence of a topological analysis of covering spaces is best described by the category theoretic approach.

## 5. Category Theoretic Modeling and Analysis

As mentioned in the introduction, category theory[19-24] is a branch of mathematics that formalizes mathematical structures into collections of objects and morphisms (or arrows) that connect them. Category theory extracts the core of all mathematical subjects to reveal and formalize simple, yet extremely powerful patterns of thinking.[23] Mathematically, a category is defined as follows[19,20]:

**Definition 1 (Category)**
1. Objects: $A, B, C, \ldots$
2. Morphisms: $f, g, h, \ldots$
3. For each morphism $f$, there are given objects, $\text{dom}(f)$ and $\text{cod}(f)$, called the domain and codomain of $f$, respectively. We write $f: A \to B$ to indicate that $A = \text{dom}(f)$ and $B = \text{cod}(f)$.
4. Given morphisms $f: A \to B$ and $g: B \to C$, there is a given morphism $g \circ f: A \to C$ called the composite of $f$ and $g$.
5. For each object $A$, there is a given morphism $1_A : A \to A$ called the identity morphism of $A$.
6. Associability $h \circ (g \circ f) = (h \circ g) \circ f$ for all $f: A \to B$, $g: B \to C$, $h: C \to D$.
7. Unit $f \circ 1_A = f = 1_B \circ f$ for all $f: A \to B$.

A category can be anything that satisfies these definitions. One of the significant features of category theory is that objects and morphisms are determined by the role they play in a category, via their relationship to other objects and morphisms, i.e., by their position in a structure, and not by what they are, or what they are made from.[20] We consider the possibility that these properties of category theory can be highly revealing when it comes to understanding decision-making problems; the category theoretic viewpoint may reveal the underlying structure. In this study, we do not examine exact mathematical formulae and proofs, as our goal is to obtain a physical insight into the single-photon decision maker via category theoretic viewpoints.

We start with a rudimentary picture of a decision-making problem, represented schematically in Fig. 6(a), where two objects are connected by a morphism. One object represents "***Decision***," and the other object represents "***Result***;" denoted by $P$ and $Q$, respectively. In other words, "**Decision**" and "**Result**" are the "*Initial*" and "*Final*" states of an apparent playing process, respectively. The single-photon-based decision maker and probabilistically behaving slot machines are absent in this simplified diagram.

### 5.1. *Product and coproduct*

As the first step, we include the notions of "*product*" and "*coproduct*," which are obtained from a basic category theoretic perspective.[19]

**Definition 2 (product and coproduct)**

In any category, a *product diagram* for objects $A$ and $B$ consists of an object ($S$) and morphisms $A \xleftarrow{p_1} S \xrightarrow{p_2} B$ satisfying the following. Given any diagram of the form $A \xleftarrow{x_1} Z \xrightarrow{x_2} B$, there exists a unique $u: Z \to S$, making the diagram

$$\begin{array}{c} Z \\ {}_{z_1}\swarrow \;\; \downarrow u \;\; \searrow^{z_2} \\ A \xleftarrow{p_1} S \xrightarrow{p_2} B \end{array} \tag{10}$$

commute, i.e., $z_1 = p_1 \circ u$ and $z_2 = p_2 \circ u$. $S$ is written as $A \otimes B$.

A diagram, $A \xrightarrow{p_1} T \xleftarrow{p_2} B$, is a *coproduct* of A and B, represented by $T = A \oplus B$, if for any $Z$ and $A \xrightarrow{z_1} Z \xleftarrow{z_2} B$, there is a unique $u: T \to Z$ with $u \circ q_1 = z_1$ and $u \circ q_2 = z_2$ as indicated in

$$\begin{array}{c} Z \\ {}_{z_1}\nwarrow \;\; \uparrow u \;\; \nearrow^{z_2} \\ A \xrightarrow{q_1} T \xleftarrow{q_2} B \end{array} \tag{11}$$

Here, we consider the physical meaning of product and coproduct in the case of decision making in a casino. From Eq. (11), the coproduct that characterizes the system $T$ corresponds to a composite of $A$ and $B$ with respect to the natures described by the combination of morphisms $q_1$ and $q_2$. When the coproduct is transformed by $u: T \to Z$, the nature of $Z$ is understood by the combination of $z_1$ and $z_2$. In the case of playing slot machines, the decision maker speculates the casino via a composite of "Decision" and "Result" of the betting, which is given by $Y(\mathcal{M})$ in Fig. 6(b). Based on an understanding of the system (as the coproduct), the composite system is given by $X(\mathcal{M})$ in Fig. 6(b), then discloses the next "Decision" and probable "Result." The latter is based on the meaning of product, given by Eq. (10), that the system ($S$) is disclosed into $A$ and $B$, with respect to the natures described by the combination of morphisms $p_1$ and $p_2$. When the system is transformed by $u: Z \to S$, the nature of $Z$ is understood by the combination of $z_1$ and $z_2$.

In repeating slot machine plays, the components involved in the composite system, i.e., "Decision" and "Result," adjust their state so that a higher reward sum is obtainable. Such an adjustment of the composite system corresponds to an endomorphism of the coproduct (see Eq. (9)). It should be emphasized that the nature of the entire system should be understood in terms of product and coproduct, which is a composite system, including hidden environmental entities and singularities, as discussed in Sec. 4. The observable entities are "Decision" and "Result." The category theoretic description has the potential to precisely handle these hidden attributes (involved through cohomology), as described below.

The product and coproduct, shown in Fig. 6(b) as a monoidal category, are summarized in a diagram, as shown in Fig. 6(c). The object "**Decision**" ($P$) is concerned with whether "Machine 1" or "Machine 2" is selected, whereas the object "**Result**" ($Q$) is concerned with whether the betting result is either "WIN" or "LOSE." The practical slot playing processes, as a whole, involves complicated environmental factors (as discussed above), but the essence is describable based on the structure shown in Fig. 6(c). This can be accomplished by incorporating all of the environmental conditions preceding the slot play into the product, and by incorporating the succeeding environmental conditions into the coproduct. Here, we provide the physical interpretation of each object and morphism. The product $P \otimes Q$, denoted by $X$ in Fig. 6(c), indicates the "*Casino Setting*," including the entire environmental condition of the slot plays. The morphism from the **Casino Setting** ($X$) to **Result** ($Q$) indicates that the **Casino Setting** generates the betting **Result** (WIN or LOSE), and all the environmental conditions are injected into "0" in $Q$, i.e., the equivalent class to the **Result**. In other words, although environmental conditions do affect the probabilistic attributes of slot machines, they cannot be distinguished in the **Result**. Such a structure is represented so that all environmental conditions are included in the kernel of the morphism $X \to Q$. According to category theory, the kernel can be regarded as a set consisting of the preceding objects and the morphism, in a short exact sequence. Therefore, "*all environmental conditions*" of the slot play, also referred to as "*Machine operation environment*," is defined as object $M$, which is shown in Fig. 7(a).[22] Here, the short exact sequence is represented by $0 \to M \to X \to Q \to 0$. The environmental conditions $M$ is projected to 0 in $Q$, meaning that $M$ is the kernel of the morphism $X \to Q$. At the same time, the **Casino Setting** affects the **Decision**, which is manifested by the morphism from $X$ to $P$. Here, the angle $\theta/2$, which was introduced in the geometrical analysis discussed earlier, is marked in the vicinity of $X$ because the **Casino Setting** involves the hidden adjustment concerned with the GOOD/BAD machine.

The coproduct $P \oplus Q$, denoted by $Y$ in Fig. 6(b), is introduced based on the representative descriptions that the slot play processes are synthesized as the combination of the **Decision** ($P$) and **Result** ($Q$). This is required to extract the knowledge needed to make better decisions in subsequent trials; i.e., the information on whether the GOOD machine is associated with either "Machine 1" or "Machine 2" based on the betting result. The coproduct $P \oplus Q$ involves all the environmental conditions after a slot play is completed. The visible portion of the coproduct ($Y$) is physically accommodated in the polarization updating strategy based on the result, so $Y$ is referred to as the "*Polarizer Setting*" ($\phi/2$) for decision making. For this setting, the hidden parameter of the slot machines $\theta/2$ is inferred. Because the morphism $Q \to Y$ dominates the polarization of the single photon source that determines the quantum mechanical probability distribution in the decision maker, the environmental conditions for the optical fields correspond to a co-kernel of the morphism $Q \to Y$. According to category theory, the co-kernel can be regarded as a set consisting of the subsequent object and morphism, in a short exact sequence. Therefore, the optical environmental conditions (namely, the "*Photon environment*") of the single-photon decision maker is placed as object $F$, as shown in Fig. 7(a).[22] Such dependencies are clearly supported by the reconfiguration strategy of the single-photon decision maker. It should be noted that all the unobservable environmental conditions after single slot-play are implicitly included in the **Photon environment** ($F$). Such a high reception of objects and morphisms reveals the expressive power of category theory.

It is noteworthy that the category theoretic picture shown in Fig. 6(c) indicates that the product $X = P \otimes Q$ corresponds to the "governor" that dominates the initial and final states of apparent slot-playing events, whereas the coproduct $Y = P \oplus Q$ is the "observer" who attempts to speculate the intension of the governor. In the mathematical context of category theory, the diagram shown in Fig. 6(c) corresponds to a representative description of the slot playing process $P \to Q$, based on the representative morphisms $X \to P$ and $Q \to Y$, belonging to a multiplicative system of morphisms. In other words, one can describe the slot playing process in a different (but equivalent) manner, based on any set of representative morphisms belonging to the multiplicative system.[22] The morphisms $X \to Q$ and $P \to Y$ are referred to as the right and left quotient morphisms of $P \to Q$, respectively, based on the representatives $X \to P$ and $Q \to Y$. The most important feature of the diagram shown in Fig. 6(c) resides in the commutative relation between $X \to P \to Y$ and $X \to Q \to Y$, which leads us to the description of the polarization updating strategy in the following section.

### 5.2. *Complex and characteristic arrow*

We can assume that the category theoretic graph shown in Fig. 6(c) indicates the relationships between objects when they have established certain stationary states, i.e., a single slot-playing process is completed, which includes adjustment of the single-photon decision maker for subsequent slot-playing. Then, we can proceed to the next slot-play, based on the prepared optical environment. In order to construct the category theoretic picture of the polarization updating strategy of the single-photon decision maker, we need to introduce the notion of "complex."

**Definition 3 (complex)**

A complex ($A^\bullet$) is a sequence of objects $\{A^j\}_{j \in Z}$ and morphisms $d_A^j : A^j \to A^{j+1}$, such that $d_A^j \circ d_A^{j-1} = 0$ for all $j$. In other words, a complex is the nature of the boundary operator, or differential operator.

Due to this nature, the object $A^{j-1}$ (in the complex) is injected as the image of the morphism $d_A^{j-1}$, into the kernel of the morphism $d_A^j$. It is noted that the quotient of the kernel of $d_A^j$, divided by the equivalent class of the image $d_A^{j-1}$, is referred to as the $j$-th order cohomology $H^j(A^\bullet)$ of the complex $A^\bullet$. One remarkable feature is that the cohomology is irrelevant to the preceding object, and is transferred to the subsequent object as the equivalent class of the zero object. In other words, the cohomology $H^j(A^\bullet)$ represents the local feature, added only to object $A^j$ in complex $A^\bullet$. Therefore, a complex describes a sequential evolution of objects, with a history of sequential cohomology addition.

It is useful to introduce "shift" or "translation" of a complex; $C^\bullet = A^\bullet[1]$ consisting of $\{C^j\}_{j \in Z} = A^{j+1}$ with $d_C^j = -1 d_A^j$.[22]

A morphism of the complex $f: A^\bullet \to B^\bullet$, is a set of morphisms $f^j: A^j \to B^j$, which satisfies $f^{j+1} \circ d_A^j = d_B^j \circ f^j$ for all $j$. Here, one of the most significant features in category theory is concerned with the "chain-wise exact sequence of complex" given by $0 \to P^\bullet \to Q^\bullet \to R^\bullet \to 0$, which consists of short exact sequences $0 \to P^j \to Q^j \to R^j \to 0$, for all $j$. It is important that the "chain-wise exact sequence of complex" induces the long exact sequence of cohomology[22];

$$\cdots \to \bullet \to \bullet \to H^{j-1}(R) \to H^j(P) \to H^j(Q) \to H^j(R) \to H^{j+1}(P) \to \bullet \to \bullet \to \cdots. \quad (12)$$

Moreover, category theory tells us that within a certain equivalent class of cohomology, one can find a "characteristic arrow," also called a "translation morphism" $R^\bullet = P^\bullet[1]$, which maintains the long exact sequence of cohomology. That is, the evolution of the "chain-wise exact sequence of complex" is described by a "triangular" structure, $P^\bullet \to Q^\bullet \to R^\bullet \to P^\bullet[1]$.[22]

### 5.3. *Octahedral structure in decision making*

Based on the category theoretical descriptions of the step-wise evolving relationship of objects, in terms of the complexes and morphisms discussed above, we can transform the single slot playing process with the single-photon decision maker (described in Fig. 7(a)), into the diagram of complexes evolving under the polarization updating strategy (as shown in Fig. 7(b)). In this diagram, the updating processes of the entire environmental condition (*M*) and the photon environment (*F*) have been introduced with the characteristic arrows indicated by the wavy lines. The positions of *M*, *F*, and the characteristic arrows, are determined based on the commutative relation between $X^\bullet \to P^\bullet \to Y^\bullet$ and $X^\bullet \to Q^\bullet \to Y^\bullet$. Because the step-wise short exact sequences $0 \to M \to X \to Q \to 0$ and $0 \to Q \to Y \to F \to 0$ fulfil certain equilibrium conditions for each slot playing process, we can derive the composite morphisms $M^\bullet \to Q^\bullet$ and $Q^\bullet \to F^\bullet$. Accordingly, the category theoretic picture in Fig. 7(b) includes triangular structures given by $M^\bullet \to Q^\bullet \to Y^\bullet \to M^\bullet[1]$ (marked by "B3") and $X^\bullet \to Q^\bullet \to F^\bullet \to X^\bullet[1]$ ("B1") corresponding to the polarization updating strategy. This diagram coincides with a physical interpretation of subsequent machine selection based on the optical environmental conditions. This is accomplished through the single-photon decision maker, which takes into consideration all environmental conditions after a slot play, to prepare the subsequent environmental conditions of slot play.

Because the diagram in Fig. 7(b) has the same complexes at the top and bottom ($M^\bullet$ and $F^\bullet$), we can construct an equivalent diagram while having $P^\bullet$ and $Q^\bullet$ placed at the top and bottom, as shown in Fig. 7(c). This indicates the commutativity of $Y^\bullet \to F^\bullet \to X^\bullet$ and $Y^\bullet \to M^\bullet \to X^\bullet$, including characteristic arrows. We can complete the diagram by adding the composite morphisms $F^\bullet \to P^\bullet$ and $P^\bullet \to M^\bullet$ as characteristic arrows, as shown in Fig. 7(d). Here, we can find another two triangular structures, given by $P^\bullet \to Y^\bullet \to F^\bullet \to P^\bullet[1]$ (denoted by "B4" in Fig. 7(d)) and $M^\bullet \to X^\bullet \to P^\bullet \to M^\bullet[1]$ ("B2") corresponding to the polarization updating strategy.

The diagrams in Fig. 7(d) present a compact picture of an octahedral structure, by directly connecting the same complexes at the top and bottom of the diagram with each other, leading to the three-dimensional diagram shown in Fig. 7(e). This structure corresponds to one of the most important consequences of "triangulated category" or "derived category", called the octahedron axiom.[19] The octahedron consists of four short exact sequences corresponding to the triangular category, and four triangular diagrams as indicated in Fig. 7(e). The triangles B1 and B2 are located in the upper half of the octahedron, while those of B3 and B4 are located in the lower half.

It is important to comment on the interpretation of the induced long exact sequence of cohomology, or the triangular structure of the complex in decision making. It is natural to associate the cohomology induced in each complex with the "local environment." For example, the betting results of each play are determined based on the spontaneous symmetry break occurring in the slot machine, which is included in the cohomology of the machine operation environment ($M^\bullet$). Generally, descriptions of the intention, will, or preference, of both the decision maker and the slot machine, are represented in cohomology. However, the short exact sequences, which involve no cohomology, restrict the evolution of complexes in the triangulated category. Therefore, the sequence of cohomology indicates the "history" of evolution experienced by each object, via the triangulated structure.

In the category theoretic context, the octahedral structure is known to be resolved into two Mayer-Vietoris sequences[22]

$$X^\bullet \to Q^\bullet \oplus P^\bullet \to Y^\bullet \to X^\bullet[1], \tag{13}$$

$$X^\bullet \to Y^\bullet \to M^\bullet[1] \oplus F^\bullet \to X^\bullet[1] \tag{14}$$

which correspond, respectively, to the two commutative diagrams shown in Figs 7(b) and 7(c). These Mayer–Vietoris sequences imply that the structure of the initially unknown structure ($X^\bullet$) is transferred to the observer ($Y^\bullet$), indicating that correct decision making is realized. In the following section, we investigate the geometrical properties based on the braid structure of the octahedral diagram.

One remark we have is that category theory provides us with a useful framework to picture complicated physical systems and their functionalities without necessarily identifying the exact physical entities. This is because category theory allows us to describe and analyze relationships and functionalities of conceptualized objects, and to derive substances of the systems and their functionalities. Therefore, category theory is especially useful for describing systems that exert functionalities in non-equilibrium open systems, involving a certain group or hierarchy of environmental systems. Furthermore, by using "functor" in category theory, functional structures can be transferred from one category to another, while preserving essential substances.

### 5.4. *Topological understanding of the decision-making strategy*

It is useful to describe the decision-making strategy employed in the photon-based decision maker (Sec. 2), in terms of the topological structures introduced in Sec. 4. As discussed in Sec. 4 and Fig. 5, the strategy is used to identify the singularity induced by a casino that generates multivaluedness in the "WIN" (and "LOSE") states in the monoidal category under study. We can represent such a singularity as a twist in a string diagram. Suppose that the casino sets "Machine 1" as the "GOOD" machine, as shown in Fig. 8(a,i). Such a singularity is represented by the twisted string shown in Fig. 8(a,ii). This is referred to as a "casino string." Fig. 8(b,i) represents another situation where "Machine 2" is set as the "GOOD" machine, in which the casino string is given by Fig. 8(b,ii). Depending on the intention of the casino, one of the alternative twists is chosen.

The player can only see the betting results, without directly observing the topological structure of the casino. Let us consider another string diagram for the decision-making process, which we call "player string." Depending on the intrinsic fluctuations of the photon environment, the machine operation environment, and the polarizer setting, the decision maker infers which of the slot machines is "GOOD." Suppose that the player speculates that "Machine 1" is a "GOOD" machine, as schematically shown in Fig. 8(a,i). In a similar way to the casino string, such a diagram is abstracted to a twisted string, as depicted in Fig. 8(a,ii).

The decision-making strategy is used to couple the casino string and the player string, as shown in Fig. 8(a,iii), followed by a connection of the two, as shown in Fig. 8(a,iv). If the player string corresponds to an adjoint (with respect to the casino string), the coupled string is successfully unfolded, as shown in Fig. 8(a,v). When the player string does not meet the casino string, as shown in Fig. 8(b), the coupled string is not unfolded (Fig. 8(b,v)).

At the same time, when the betting result is "LOSE," the player speculates that the player string may not be the adjoint of the true casino string. Hence, an adjustment of the polarizer setting is performed, which is an endomorphism of the object ($Y$). Such a picture, involving unfolding the knots, is numerically demonstrated in Sec. 5.5 on the basis of the octahedron structure.

Here, it should be noted that there exists an important criterion for successful decision making. This criterion is explained by the notion of "section." Consider the short exact sequence $0 \to F \to X \to P \to 0$. This indicates that the "photon environment"($F$), affects the "decision" ($P$). If a reversal homomorphism ($s: P \to X$) exists for homomorphism $f: X \to P$, such that it satisfies $f \circ s = 1$, $s$ is called a section of homomorphism $f$. Further, the reversal short exact sequence $0 \to P \to X \to F \to 0$ is derived, and a quotient object ($X / P$) is well-defined.

In the photon-based decision maker, the state of a single photon is inferred from the result of polarization observation, owing to the particle nature of a single photon. In turn, the sequence $0 \to M \to X \to Q \to 0$ corresponds to provide the betting results. The existence of a section to the homomorphism $g: X \to Q$, ensures that the casino (more specifically, the machine operation environment) is classified well by the betting results. However, when too large a fluctuation is present in casino, one cannot identify the section morphism to $g$. Hence, an inference regarding the state of the casino has no meaning. Indeed, the reward probability difference between "GOOD" and "BAD" becomes subtle, the derivation of section morphism is very difficult, and the photon-based decision strategy does not work well. We consider that deeper insight into the criterion, on the basis of section morphism sectors in quotient objects, would be an important factor in future studies.

**5.5. *Braid structure in decision making***

In order to deepen the understanding of the physical and mathematical implications of single-photon-based decision making, via geometrical considerations, we have simulated the evolution of the polarization updating process. This is based on its "braid structure," of the octahedron shown in Fig. 7(e) regarding the relation between the hidden variable, $\theta / 2$, and polarization setting, $\phi / 2$.

As a series of decision making and polarization updating processes, the diagram shown in Fig. 7(d) can be extended by appending shifted diagrams of octahedral structures, as shown in Fig. 9(a). More specifically, the extended diagram in Fig. 9(a) is derived by repeating the diagram in Fig. 7(d), while swapping the positions of $P^{\bullet}$ and $Q^{\bullet}$. Consequently, the four short exact sequences are arranged sequentially. Following the category theoretic context, this diagram produces an octahedral "braid structure"[22,27], shown in Fig. 9(b), which consists of the following four exact sequences as braids;

$$
\begin{array}{ccc}
\text{Braid1} & \text{Braid4} & \\
0 & 0 & \\
\downarrow & \downarrow & \\
0 \to M^{\bullet} \to X^{\bullet} \to P^{\bullet} \to 0 & \text{Braid2} \\
\downarrow \quad \downarrow \quad \downarrow & \\
0 \to M^{\bullet} \to Q^{\bullet} \to Y^{\bullet} \to 0 & \text{Braid3} \\
\downarrow \quad \downarrow & \\
F^{\bullet} \to F^{\bullet} & \\
\downarrow \quad \downarrow & \\
0 \quad 0 &
\end{array}
\tag{15}
$$

This braid structure reveals the geometrical structure, or interdependence underlying the single-photon decision maker, in a totally simplified manner. Here, we further explore the braid concept by examining the "knots" of the braids. Because our main concern in decision making is whether Machine 1 or 2 corresponds to the GOOD machine, we can analyze the relative behaviors of the braids, with respect to the hidden variable of the Casino Setting ($\theta/2$), as follows.

**[Study 1: Knots at *X*]** Braids **1** and **2** intersect each other at *X*. Because *X* represents the Casino Setting ($\theta/2$), let us assume that $\theta > 0$ means that **Braid 1** stays over **Braid 2**, whereas $\theta < 0$ represents the converse. If the **Casino Setting** is unchanged (i.e., $\theta/2$ is constant), one braid is always on top of the other, and hence there is no knotting of braids. This scenario is schematically shown in Fig. 9(c).

**[Study 2: Knots at *Y*]** Braids **3** and **4** intersect at *Y*, which physically corresponds to the **Polarizer Setting** ($\phi/2$). A knot is induced when two braids are "wrapped," as schematically shown in Fig. 9(d). The purpose of the decision maker is to unfold the knots at *Y*, such that an adequate $\phi/2$ is derived by inferring the **Casino Setting** ($\theta/2$). When quantitatively analyzing the braids, we investigate the system simulated in the numerical analysis shown earlier. We let the reward probabilities of slot machines 1 and 2 be 0.6 and 0.4, respectively, and all other conditions of numerical simulation are the same as previously discussed. Fig. 9(e) shows an incidence histogram of the number of knots at *Y*. As time elapses, knots are unfolded, meaning that the **Casino Setting** ($\theta/2$) is inferred by the decision maker. Fig. 9(f) evaluates when the "last" knot in 500 consecutive plays is induced in the system; the incidence frequency decays quickly, meaning that polarization adaptation is completed promptly.

**[Study 3: Knots at *P*]** Braids **2** and **4** intersect at *P*, which physically corresponds to the **Decision**. The machine selection is probabilistically conducted based on the polarization ($\phi/2$) of an incident single photon. This means that the knot of **Machine Selection** (*P*) is not equivalent to that of the **Polarizer Setting** (*Y*). As shown in Fig. 9(g), the histogram of the total number of knots at *P* differs from that of the **Polarizer Setting** (Fig. 9(e)), as the number of knots at *P* is larger than at *Y*. As shown in Fig. 9(h), which displays the last appearance of knots over 500 plays, knots are induced, even after hundreds of plays are conducted. These observations clearly establish the fundamental architecture of the decision-making problem, and the dynamics of the single-photon decision maker.

**[Study 4: Knots at *F*]** Braids **1** and **4** intersect at *F*, which corresponds to the polarization of a single photon. This polarization is deterministically configured by the **Polarizer Setting**, with the properties of knots at *F* equal to those at *Y* (see **Study 2**).

**[Study 5: Knots at *Q* and *M*]** Braids **1** and **3** intersect at *Q*, and **Braids 2** and **3** intersect at *M*. The physical interpretation of *Q* is the WIN/LOSE betting results, and *M* is the **Machine operation environment**. Neither of these entities can be controlled by the decision maker, hence knots at *Q* and *M* are never unfolded.

This can be understood by considering the case where even completely correct decisions cannot avoid "LOSE" events, if the reward probability of the GOOD slot machine is not 100%.

Furthermore, both $Q$ and $M$ are on the same braid, **Braid 3**: $\to M^{\bullet} \to Q^{\bullet} \to Y^{\bullet} \to$, so that the other object on **Braid 3** ($Y$) physically indicates that the polarized angle exhibits fluctuations due to the uncontrollable entities on **Braid 3**. This is another insight provided by the category theoretic analysis.

Before the conclusion, we make a few remarks regarding extensions of the present study, and about its relevance to practical decision-making problems. First, this study uses the simplest case (two-armed bandit problem), hence the number of singularities is essentially only one (which of Machine 1 or Machine 2 has the higher reward probability). When the number of singularities increases (e.g., for an increased number of slot machines), the problem becomes complex and difficult. For example, in case of a four-armed bandit, which has been successfully resolved experimentally using single photons,[29] the number of singularities becomes six. Such theoretical extensions are interesting and important for future studies. Meanwhile, this study summarizes that the environmental fluctuations included in the objecsts $M$ and $F$, affect the decision making. Recently, Naruse *et al.* experimentally demonstrated that a chaotically-oscillated time-series (generated by lasers) performed better than pseudorandom numbers, including colored noise.[30] Category theoretical approach is an interesting area of future study for such environmental and critical entities.

Furthermore, in our study there is only one decision maker. When the number of decision makers increases, the problem is called a competitive multi-armed bandit problem, where a conflict of decisions may occur. Such a conflict is difficult to resolve if the situation involves a Nash equilibrium.[31] Kim *et al.* developed a physical architecture to maximize the total reward of all players by introducing inter-decision-maker dynamics.[31] The categorical understanding of this will be explored in a future study. In this context, group-decision making and group consensus have been extensively studied.[32,33] The category theoretical implications of this research would also be interesting topics for future studies.

## 6. Conclusion

We have presented a theoretical analysis of single-photon-based decision making using category theory. We also introduced the concept of DECISION and WIN/LOSE circles, through which the evolution of polarization adaptation was theoretically formulated and analyzed. The simulation and experimental results agreed, in accordance with the geometrical modeling. We generalized this approach using category theory. Beginning with the topological picture of a decision-making problem, we demonstrated that the octahedral structure in a triangulated category clearly reveals the underlying mechanisms of the single-photon-based decision-making process. The effects of the environment were appropriately included as objects in the octahedral structure. The Mayer–Vietoris sequences also support the principle of decision making, in the form of transferring the initially unknown structure to the observer. A topological picture of the decision-making strategy, with the concept of twisted strings, also helps to understand the underling architecture. The braid structure, used in the triangulated category, provided various insights. These included quantitative decision-making metrics, such as the fact that the braid knots unfolding corresponds to an adaptation to a given problem, and that braids that cannot be unfolded represent uncontrollable entities that are inherent to the system. This study provides a fundamental analysis of photon-based decision making, and paves the way for future use of the category theoretic approach, for the general understanding and design of intelligent machines.

**Acknowledgments**

The authors would like to thank M. Agu and K. Kitahara, for their encouragement toward the study of functionalities in a non-equilibrium open system. This work was partly supported by the Grant-in-aid in Scientific Research (A) (JP17H01277) and the Core-to-Core Program, A. Advanced Research Networks from the Japan Society for the Promotion of Science, CREST program (JPMJCR17N2) from by Japan

Science and Technology Agency, and Agence Nationale de la Recherche, France, through the SINPHONIE and PLACORE project (Grant No. ANR-12-NANO-0019 and ANR-13-BS10-0007). M.N. is supported by the Université Grenoble Alpes, France, through an invited professorship.

**References**

1. N. Daw, J. O'Doherty, P. Dayan, B. Seymour and R. Dolan, Cortical substrates for exploratory decisions in humans, *Nature* **441** (2006) 876–879.
2. R. S. Sutton and A. G. Barto, *Reinforcement Learning: An Introduction* (The MIT Press, Cambridge, 1998).
3. M. Naruse, M. Berthel, A. Drezet, S. Huant, M. Aono, H. Hori and S. -J. Kim, Single-photon decision maker, *Scientific Reports* **5** (2015) 13253.
4. L. Lai, H. Gamal, H. Jiang and V. Poor, Cognitive medium access: exploration, exploitation, and competition, *IEEE Transactions on Mobile Computing* **10** (2011) 239–253.
5. S. -J. Kim and M. Aono, Amoeba-inspired algorithm for cognitive medium access, *NOLTA* **5** (2014) 198–209.
6. D. Agarwal, B. -C. Chen and P. Elango, Explore/exploit schemes for web content optimization, In *Proc. IEEE Int. Conf. Data Mining* (2009), pp. 1–10.
7. L. Kocsis and C. Szepesvári, Bandit based Monte Carlo planning, In *European Conf. Machine Learning* (2006), pp. 282–293.
8. W. H. Press, Bandit solutions provide unified ethical models for randomized clinical trials and comparative effectiveness research, *Proceedings of the National Academy of Sciences* **106** (2009) 22387–22392.
9. H. Takahashi, Molecular neuroimaging of emotional decision-making, *Neuroscience Research* **75** (2013) 269–274.
10. E. M. Pothos and J. Busemeyer, A quantum probability explanation for violations of 'rational' decision theory, *Proceedings of the Royal Society of London B: Biological Sciences* (2009) rspb-2009.
11. T. Cheon and T. Takahashi, Interference and inequality in quantum decision theory, *Physics Letters A* **375** (2010) 100–104.
12. E. Haven and A. Khrennikov, A, Quantum probability and the mathematical modelling of decision-making, Phil. Trans. A **374** (2016) 2058.
13. P. Auer, N. Cesa-Bianchi and P. Fischer, Finite-time analysis of the multi-armed bandit problem, *Machine Learning* **47** (2002) 235–256.
14. S. -J. Kim, M. Aono and M. Hara, Tug-of-war model for the two-bandit problem: Nonlocally-correlated parallel exploration via resource conservation, BioSystems **101** (2010) 29–36.
15. S. -J. Kim, M. Aono, M. Hara, Tug-of-war model for multi-armed bandit problem, *Lecture Notes on Computer Science* **6079** (2010) 69–80.
16. S. -J. Kim, M. Aono and E. Nameda, Efficient decision-making by volume-conserving physical object, *New Journal of Physics* **17** (2015) 083023.
17. S. -J. Kim, M. Naruse, M. Aono, M. Ohtsu and M. Hara, Decision maker based on nanoscale photo-excitation transfer, *Scientific Reports* **3** (2013) 2370.
18. M. Naruse, W. Nomura, M. Aono, M. Ohtsu, Y. Sonnefraud, A. Drezet, S. Huant and S. -J. Kim, Decision making based on optical excitation transfer via near-field interactions between quantum dots, *Journal of Applied Physics* **116** (2014) 154303.
19. S. Mac Lane, *Categories for the Working Mathematician* (Springer, Berlin, 1971).
20. S. Awodey, *Category Theory* (Oxford University Press, Oxford, 2010).
21. H. Simmons, *An Introduction to Category Theory* (Cambridge University Press, Cambridge, 2011).
22. B. Iversen, *Cohomology of sheaves* (Springer-Verlag, Berlin, 1986).
23. D. I. Spivak, *Category Theories for the Sciences* (The MIT press, Cambridge, 2014).
24. M. Kashiwara, P. Schapira, *Categories and Sheaves* (Springer, Berlin, 2006).
25. A. Beveratos, R. Brouri, T. Gacoin, J. -P. Poizat, P. Grangier, Nonclassical radiation from diamond nanocrystals, *Physical Review A* **64** (2001) 061802.
26. L. Rondin, G. Dantelle, A. Slablab, F. Grosshans, F. Treussart, P. Bergonzo, S. Perruchas, T. Gacoin, M. Chaigneau, H. -C. Chang, V. Jacques, J. -F. Roch, Surface-induced charge state conversion of nitrogen-vacancy defects in nanodiamonds, *Physical Review B* **82** (2010) 115449.
27. M. Berthel, O. Mollet, G. Dantelle, T. Gacoin, S. Huant, A. Drezet, Photophysics of single nitrogen-vacancy centers in diamond nanocrystals, *Physical Review B* **91** (2015) 035308.
28. B. Iversen, Octahedra and braids, *Bulletin de la Société Mathématique de France* **114** (1986) 197–213.

29. M. Naruse, M. Berthel, A. Drezet, S. Huant, H. Hori, S.-J. Kim, Single photon in hierarchical architecture for physical decision making: Photon intelligence, *ACS Photonics* **3** (2016) 2505–2514.
30. M. Naruse, Y. Terashima, A. Uchida, S. -J. Kim, Ultrafast photonic reinforcement learning based on laser chaos, *Scientific Reports* **7** (2017) 8772.
31. S.-J. Kim, M. Naruse, M. Aono, Harnessing the computational power of fluids for optimization of collective decision making, *Philosophies* **1** (2016) 245-260.
32. G. Li, G. Kou, Y. Peng, A group decision making model for integrating heterogeneous information, *IEEE Transactions on Systems, Man, and Cybernetics: Systems* (2016). http://dx.doi.org/10.1109/TSMC.2016.2627050
33. T. L. Saaty, Decision making with the analytic hierarchy process, *International Journal of Services Sciences* **1** (2008) 83-98.

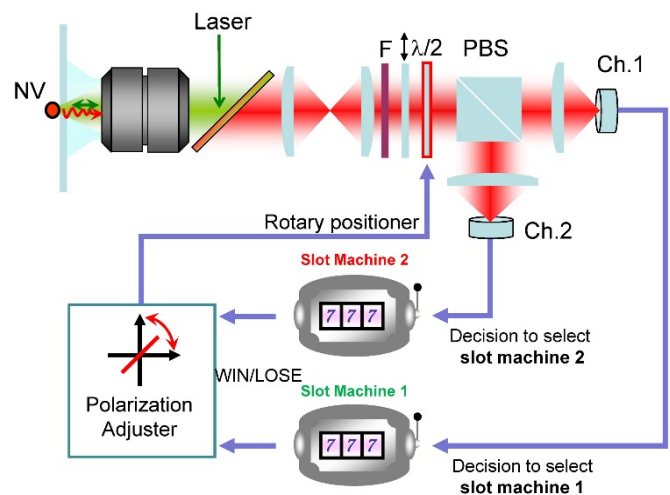

Fig. 1. Architecture of a single-photon decision maker. The polarization of single photons is configured such that the higher-reward-probability slot machine is selected. See reference (3) for details. This study is concerned with the theoretical background of decision making. (Adapted by permission from Macmillan Publishers Ltd: Scientific Reports [3], Copyright 2015)

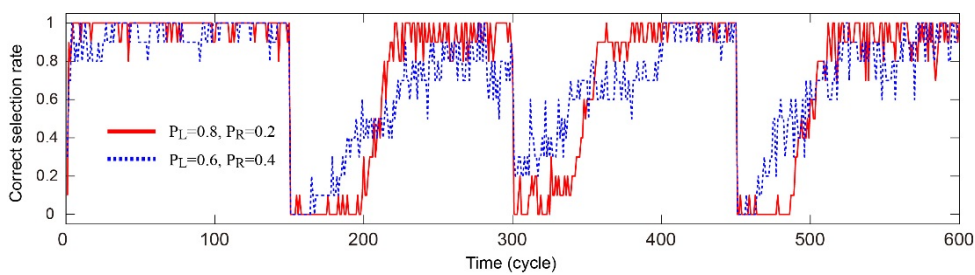

Fig. 2. Experimental demonstration of a single-photon decision maker. The reward probabilities of slot machines 1 and 2 are configured as {0.8 and 0.2} (solid curve) and {0.6 and 0.4} (dotted curve). These situations, including experimental results, are investigated in the theoretical modeling and analysis. (Adapted by permission from Macmillan Publishers Ltd: Scientific Reports [3], Copyright 2015)

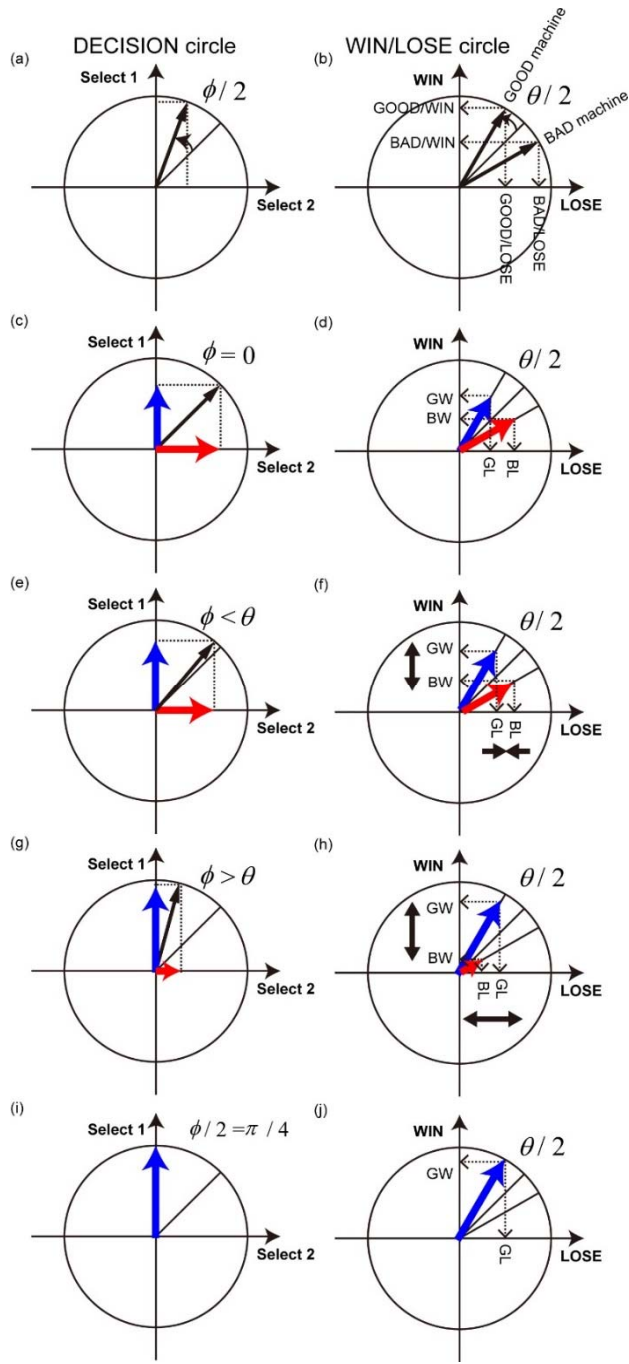

Fig. 3. Geometry-based modeling of a single-photon decision maker. Two circles are introduced: (a) *DECISION circle* where the angle $\phi$ indicates the polarization of a single photon and (b) *Win/Lose circle* where the angle $\theta$ indicates the GOOD and BAD slot machines. The decision given, based on the Decision circle, is mapped onto this Win/Lose circle. (c–j) Explaining the mechanism behind the single-photon decision maker autonomously shifting towards an accurate decision. (c, d) Initial situation ($\phi = 0$). (e, f) Second stage ($\phi < \theta$). (g, h) Third stage ($\phi > \theta$). (i, j) Final stage ($\phi/2 = \pi/4$). See main text for details.

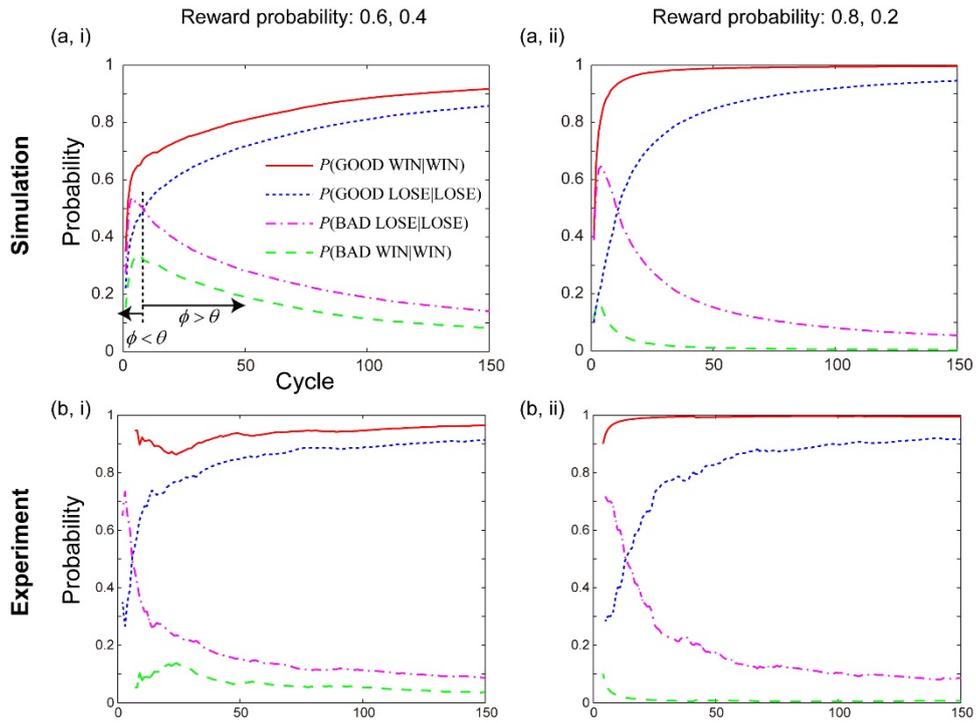

Fig. 4. Analysis of simulations and experiments from the geometry-based modeling aspect. Conditional probabilities of *P*(GOOD WIN | WIN), *P*(BAD WIN | WIN), *P*(GOOD LOSE | LOSE) and *P*(BAD LOSE | LOSE) are respectively depicted by the solid red, dashed green, dotted blue, and dash-dot magenta curves. The reward probabilities are (i) 0.6 and 0.4 (ii) 0.8 and 0.2. The simulation (a), experiment (b), and theoretical predictions are in agreement.

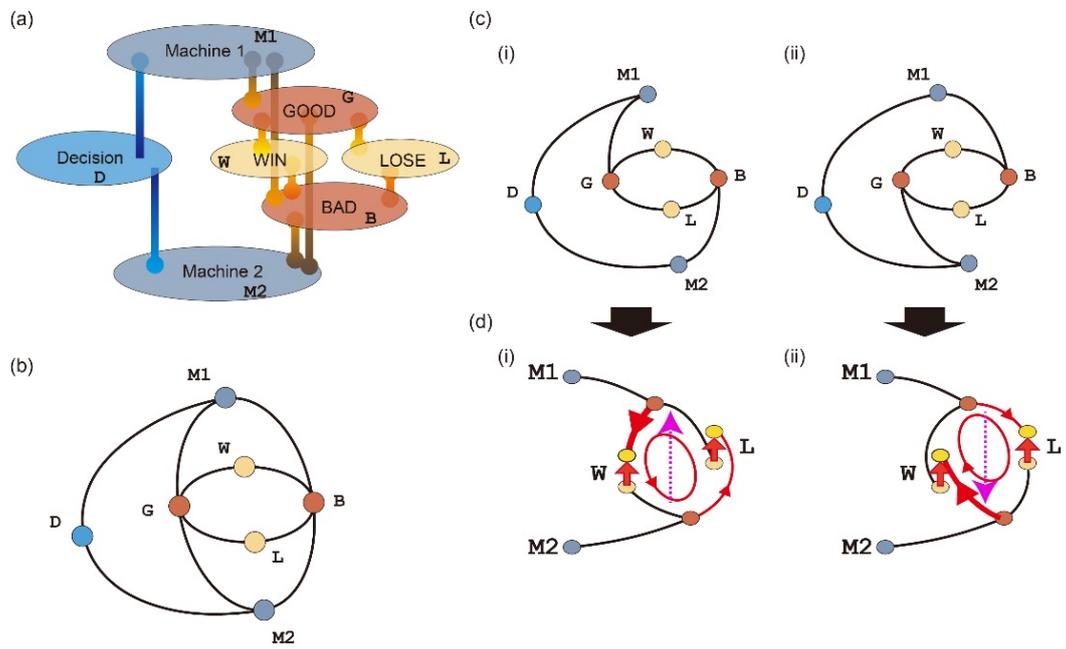

Fig. 5. (a) A schematic representation of the covering space of the two-armed bandit problem. The components of the subject matter are related to each other. (b) The covering space of (a) is abstracted as a diagram. (c) Depending on the intension of casino, *singularity* emerges. Fig. (c,i) and (c,ii) respectively show diagrams when Machine 1 (`M1`) and Machine 2 (`M2`) is assigned to the GOOD machine. (d) With such a singularity, *multivaluedness* appears in the Win and Lose nodes; in the case of Fig. (d,i), winning is more likely to be achieved via `M1` rather than `M2`.

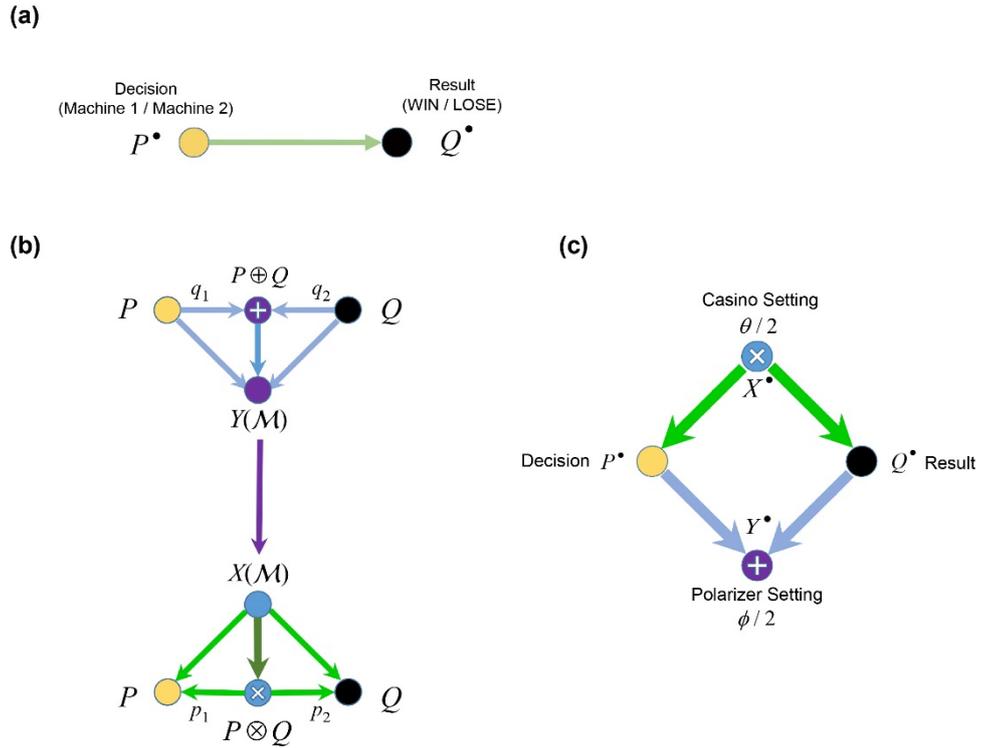

Fig. 6. Category theoretic modeling (I). (a) Rudimentary picture of decision making; Decision (*P*) directly relates to Result (*Q*). (b) Product and coproduct, involving Decision and Result. The system evolves as a monoidal category. (c) Product and coproduct are associated with "Casino Setting" and "Polarizer Setting" in the detailed modeling.

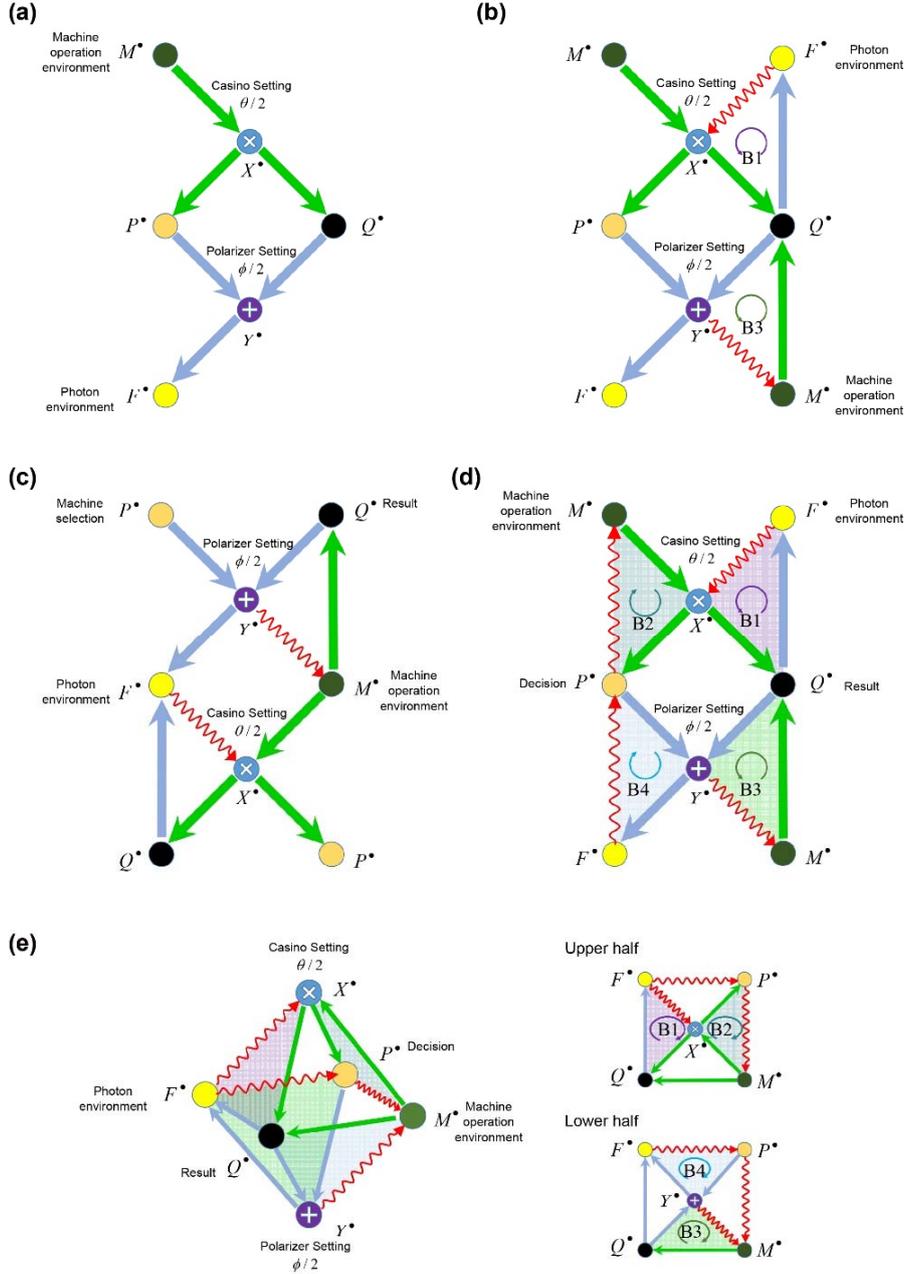

Fig. 7. Category theoretic modeling (II). (a-d) Synthesis of octahedron structure. (a) Addition of two important objects (complexes): Machine operation environment ($M^\bullet$). Photon environment ($F^\bullet$) (b) Addition of two composite morphisms ($M^\bullet \to Q^\bullet$ and $Q^\bullet \to F^\bullet$) and two translation morphisms ($Y^\bullet \to M^\bullet[1]$ and $F^\bullet \to X^\bullet[1]$). (c) Another commutative diagram ($Y^\bullet \to F^\bullet \to X^\bullet$ and $Y^\bullet \to M^\bullet \to X^\bullet$). (d) Addition of two translation morphisms ($P^\bullet \to M^\bullet[1]$ and $F^\bullet \to P^\bullet[1]$) and four triangulated structures (B1, …, B4) are derived. (e) Octahedral structure. The triangles B1 and B2 are located in the upper half, and those of B3 and B4 are in the lower half.

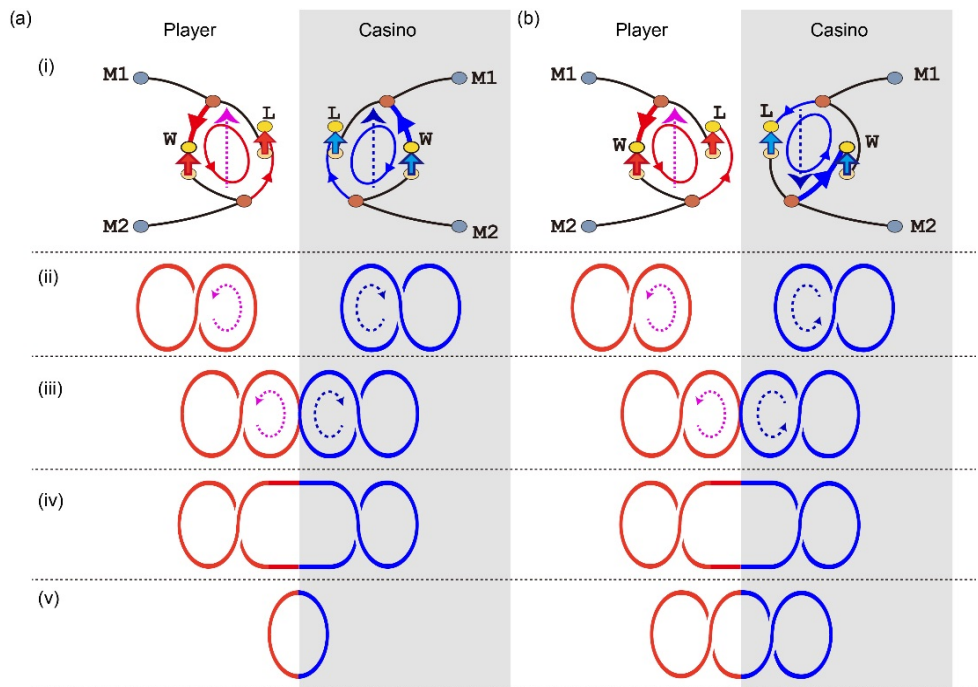

Fig. 8. Topological understanding of the photon-based decision-making strategy. (i,ii) The singularity of the casino is represented by a twisted string (casino string), while the decision of the player is also shown by a twisted string (player string). The singularity of the casino has different polarity in (a) and (b), where (a) shows the case when Machine 1 is a GOOD machine, whereas (b) indicates the opposite. (iii) The decision-making strategy involves coupling the casino string and the player string. The coupled string, shown in (iv), is unfolded when the player string is adjoint with respect to the casino string, whereas the coupled string is not unfolded when the two are not in adjoint.

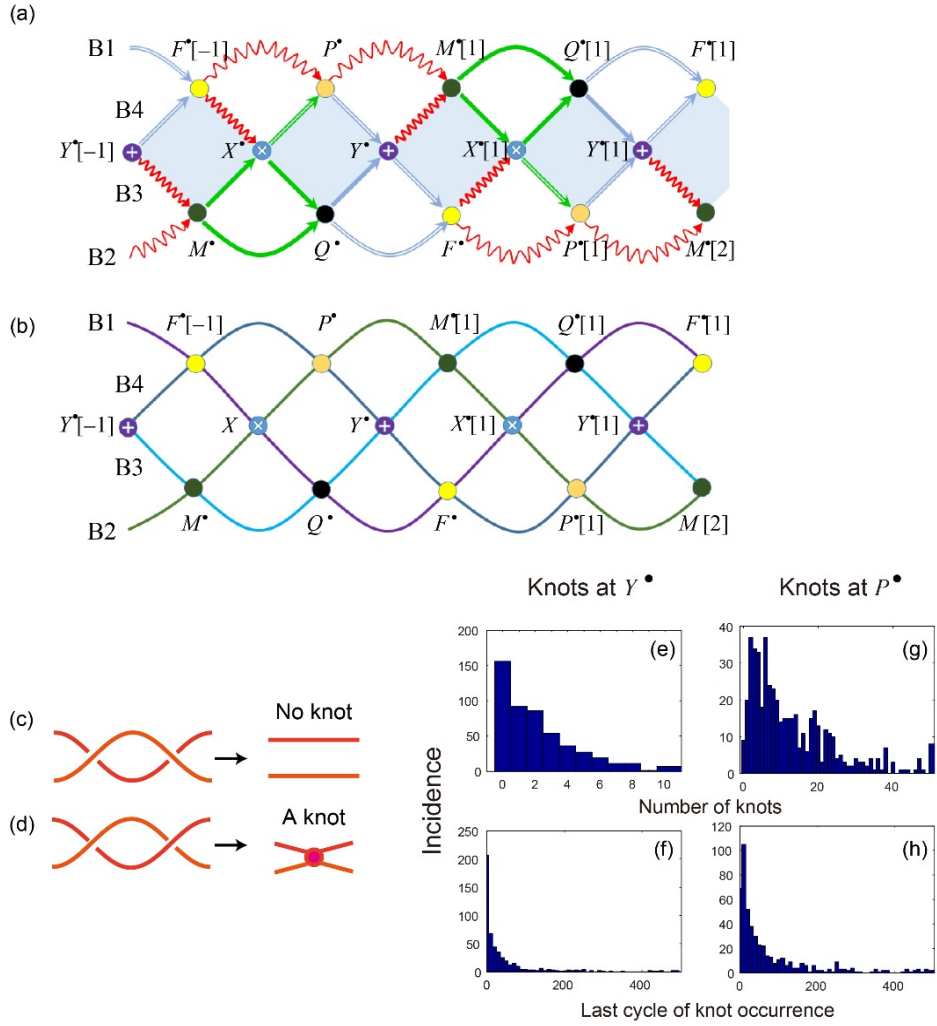

Fig. 9. Braids and knots in decision making. (a) Concatenated structure of the octahedral structure. (b) Four braids are obtained from the octahedron structure: B1: $X^\bullet \to Q^\bullet \to F^\bullet$, B2: $M^\bullet \to X^\bullet \to P^\bullet$, B3: $M^\bullet \to Q^\bullet \to Y^\bullet$, B4: $P^\bullet \to Y^\bullet \to F^\bullet$ (c, d) Knots of braids. There is no knot when one braid stays on top of the other (c), whereas a knot is induced in the situation shown in (d). (e–h) Evaluation of the knots at certain positions in the braid structure. Autonomous decision making corresponds to unwrapping the knots at the complexes of Polarizer Setting ($Y^\bullet$) and Machine selection ($P^\bullet$), which are demonstrated respectively in (e, f) and (g, h).